\documentclass[aps,prb,reprint,superscriptaddress,amsmath,amssymb,floatfix,twocolumn]{revtex4-1}
\usepackage{graphicx}
\usepackage{subfigure}
\usepackage{color}
\newcommand{\red}[1]{}

\newcommand{\beq}{\begin{equation}}
\newcommand{\eeq}{\end{equation}}

\newcommand{\tjjk}{t\!-\!J_1\!-\!J_2\!-\!K}
\newcommand{\tV}{\tilde{V}}

\begin{document}

\title{Superconducting states of a $\mathbf{J}_\mathbf{1}$-$\mathbf{J}_\mathbf{2}$-$\mathbf{K}$ model for iron pnictides}

\author{Rong Yu}
\affiliation{Department of Physics, Renmin University of China, Beijing 100872, China}
\author{Andriy H. Nevidomskyy}
\affiliation{Department of Physics and Astronomy, Rice University, Houston,
TX 77005}

\begin{abstract}
We study the symmetry and strength of the superconducting pairing in a two-orbital \mbox{$\tjjk$} model for iron pnictides using the salve boson mean-field theory. We show that the nearest-neighbor biquadratic interaction $-K (\vec{S}_i \cdot \vec{S}_j)^2$ inflences the superconducting pairing phase diagram by promoting the $d_{x^2-y^2}$ $B_{1g}$ and the $s_{x^2+y^2}$ $A_{1g}$ channels. The resulting phase diagram consists of several competing pairing channels, including an isotropic $s_{\pm}$ $A_{1g}$ channel, an anisotropic $d_{x^2-y^2}$ $B_{1g}$ channel, and two $s+id$ pairing channels. We have investigated the evolution of superconducting states with electron doping, and find that with the biquadratic interaction various pairing channels may dominate at different doping concentrations. We show that this is crucial in understanding the doping evolution of superconducting gap anisotropy observed in some angle resolved photoemission spectrum measurements.
\end{abstract}

\maketitle

\section{Introduction}
In iron pnictides, the superconductivity~ \cite{Kamihara_FeAs,Zhao_Sm1111_CPL08} emerges near
an antiferromagnetically ordered state~\cite{Cruz} in the phase diagram. This implies a
strong interplay between the superconductivity and magnetism in these materials.
It has been observed that the parent compounds of iron pnictides have a $(\pi,0)$ antiferromagnetic order, which could arise either within a weak-coupling approach invoking
a Fermi surface nesting,\cite{Graser,Ran,Knolle}
or from a strong-coupling approach whose starting point is a
local moment $J_1-J_2-K$ model.\cite{Si,Yildirim,Ma,Fang:08,Xu:08,Si_NJP,Dai,Uhrig}

In the strong-coupling approach, the $(\pi,0)$ antiferromagnetic order can arise from a $J_1-J_2$ model for sufficiently large $J_2$. But the strong anisotropic magnetic excitations in the paramagnetic phase observed from inelastic neutron scattering measurements prompts the necessity and importance of including a biquadratic exchange interaction $-K(\vec{S}_i\cdot\vec{S}_j)^2$ between the nearest-neighbor (n.n.) pair of spins.\cite{Zhao, Diallo, Harriger, Ewings, WysockiNP11, Stanek11, Yuetal12} From the theoretical perspective, the emergence of the biquadratic interaction is natural when the system contains multiple orbitals and when the correlations are strong in the metallic ground state: the low-energy effective Hamiltonian in the spin sector should consist of not only two-spin Heisenberg interactions, such as $J_1$ and $J_2$ Heisenberg exchange between n.n. and next-nearest-neighbor (n.n.n.) spins on a square lattice, but also interactions involving larger number of spins, particularly, the \red{biquadratic} interaction for spin size $S\geqslant 1$\cite{Fazekas}.

In a previous work, we have shown that the biquadratic interaction is crucial in understanding the spin dynamics in the paramagnetic phase of the parent iron pnictides.\cite{Yuetal12} Given the close correlations between magnetic excitations and the nearby superconductivity in the phase diagram, it would be equally interesting to ask how the biquadratic interaction may affect superconductivity in these materials.

The pairing symmetry of the iron-based superconductors has been studied via various experimental techniques.
Angle resolved photoemission spectroscopy (ARPES) measurements find the superconducting gap to be
nodeless and isotropic at both the hole and electron pockets in a number
of materials~\cite{Ding08,Kondo08,Liu11,Xu11,RichardDing11}. Neutron scattering measurements
on these compounds observe a clear spin resonance mode
in the superconducting state~\cite{Christianson08,ZhangDai11,Lumsden09,Chi09}.
These are consistent with an $s_{\pm}$ pairing channel, with the pairing order parameter $\Delta_{s\pm}=\Delta_0\cos(k_x)\cos(k_y)$ changing sign
between the hole pockets near the Brillouin zone (BZ) center and the electron pockets near the zone corner,
which arises within both weak-coupling
and strong-coupling approaches~\cite{Kuroki08,Mazin08,Wang08,Graser,Nicholson11,Seo08,ChenZhang09,Goswami10,Yu11}.

Recent experiments find evidences of anisotropic or even nodal superconducting gaps on several iron pnictide compounds in either underdoped or heavily doped regimes. For example, for the heavily hole doped compound KFe$_2$As$_2$, thermal conductivity and penetration-depth measurements indicate the presence of nodes,\cite{Dong10,Hashimoto10}  suggesting a different pairing symmetry to its optimally doped counterpart.\cite{Maitietal11, Maitietal12, Thomaleetal11} Experimentally, the issue is not settled, with thermal conductivity measurements interpreted as evidence for the $d$-wave symmetry of the order parameter\cite{Reid12}, consistent with predictions of the functional renormalization group~\cite{Thomaleetal11}, whereas the ARPES measurements are indicative of an extended $s$-wave state with accidental nodes.\cite{Okazaki12} A recent ARPES study of heavily hole-doped Ba$_{0.1}$K$_{0.9}$Fe$_2$As$_2$ found nodes in the small $\epsilon$ hole pockets off-centered around the $M$ point at the edge of the Brillouin zone~\cite{Xu13}.
Also in the underdoped NaFeAs and in the undoped LiFeAs, anisotropic but nodeless
superconducting gaps along the Fermi pockets have been identified.\cite{GeFeng13,UmezawaWang12,AllanDavis12} All these results question the validity of isotropic $s_{\pm}$ pairing in the underdoped and/or overdoped regimes.

In this paper, we study the superconducting states of a strong-coupling two-orbital $\tjjk$ model with electron doping using a slave-boson mean-field theory. By comparing to the results of a $t-J_1-J_2$ model, we show that a moderate biquadratic coupling $K$ changes the superconducting pairing phase diagram dramatically. It lowers the energy of both the $d_{x^2-y^2}$ $B_{1g}$ and $s_{x^2+y^2}$ $A_{1g}$ pairing channels, and thus favors an $s+id$ pairing. We are particularly interested in the doping evolution of superconductivity. Our results indicate that the biquadratic interaction is a key ingredient in tuning the relative stability of various superconducting states as a function of doping. This is in dramatic difference to the case of absent biquadratic interaction ($K=0$), where the $s_{\pm}$ pairing is found to be robust in the entire range of studied dopings. By contrast, a moderate biquadratic interaction leads to a serious change of the pairing symmetry, pairing strength, and gap anisotropy with doping, favouring $s+id$ and even pure $d$-wave pairing in the heavily electron- and hole-doped regimes. We show that this is important in understanding the different pairing states observed in recent experiments.

The remainder of the paper is organized as follows. In Section~\ref{Sec:Model}, we introduce
the two-orbital $\tjjk$ model and describe the slave-boson mean-field
theory for superconductivity used in this work. In Section~\ref{Sec:PhD} we show how the biquadratic coupling
$K$ affects the superconducting phase diagram.
In Section~\ref{Sec:Doping} we discuss how the superconducting states evolve with doping.
Further discussion in connection with existing theories and experiments is included in Section~\ref{Sec:Disc}. Section~\ref{Sec:Conclusion} contains a few concluding remarks.

\section{Model and Methods}\label{Sec:Model}
We consider a two-orbital $\tjjk$ model on a two-dimensional (2D) square lattice. It can be obtained from a two-orbital Hubbard model via a perturbative expansion in the strong coupling limit~\cite{Fazekas}.  The reason behind studying the two-orbital (as opposed to the full five-orbital) model is  the universally accepted fact that the major contribution to the Fermi surface comes
from the iron $t_{2g}$ orbitals ($d_{xz}, d_{yz}, d_{xy}$), whereas the $e_g$ orbital weight is very small\cite{Graser09, Daghofer10}. Additionally, the $d_{xz}$ and $d_{yz}$ orbitals carry most of the spectral weight\cite{Graser09} and the $d_{xy}$ orbital can thus be neglected in the first approximation. It is true that in order to obtain \emph{all} the Fermi pockets observed in ARPES, one needs to consider all 5 Fe orbitals\cite{Kuroki08}, and one of us has studied superconductivity in such a 5-orbital $t-J_1-J_2$ model~\cite{Yu14,Yu11}. However, inclusion of the biquadratic $K$ term would be highly non-trivial for such a \red{5-orbital} 
model and for reasons of transparency of the analysis, we chose to focus on the two-orbital model first, with the hope that our central results should remain valid upon inclusion of other orbitals. The detailed comparison with the \red{5-orbital} 
model will be made in Section~\ref{Sec:Disc}.
The two-orbital $\tjjk$ Hamiltonian reads as follows:
\begin{eqnarray}\label{Eq:Ham:tJ1J2K}
H=&-& \sum_{i<j,\alpha,\beta,s} t_{ij}^{\alpha \beta}
c^{\dagger}_{i\alpha s}c_{j\beta s}+h.c.
-\mu \sum_{i,\alpha}n_{i\alpha} \nonumber \\
&+&\sum_{\langle ij\rangle,\alpha,\beta} J_{1}^{\alpha \beta}
\left(\vec{S}_{i\alpha}\cdot \vec{S}_{j\beta}-\frac{1}{4}n_{i\alpha} n_{j\beta}\right)
\nonumber \\
&+&\sum_{\langle \langle ij\rangle \rangle,\alpha,\beta} J_{2}^{\alpha \beta}
\left(\vec{S}_{i\alpha}\cdot \vec{S}_{j\beta}-\frac{1}{4}n_{i\alpha} n_{j\beta}\right)\nonumber\\
&-&\;\sum_{\langle ij\rangle} K
\left[(\sum_{\alpha}\vec{S}_{i\alpha})\cdot (\sum_{\beta}\vec{S}_{j\beta})\right]^2,
\end{eqnarray}
where $c^{\dagger}_{i\alpha s}$
creates an electron at site $i$, with orbital index $\alpha$ and spin
projection $s$; $\mu$ is the chemical potential and
$t_{ij}^{\alpha \beta}$ the hopping matrix. The orbital index $\alpha=1,2$ correspond to
the iron  $3{d}_{xz}$ and $3d_{yz}$ orbitals, respectively.
The nearest-neighbor (n.n., $\langle ij\rangle$)
and next-nearest-neighbor (n.n.n., $\langle \langle ij \rangle \rangle$)
exchange interactions are respectively denoted by $J_{1}^{\alpha \beta}$ and $J_{2}^{\alpha \beta}$. $K$ is the coupling for the n.n. biquadratic interaction. The spin operator
$\vec{S}_{i\alpha}=\frac{1}{2}\sum_{s,s^{'}}c^{\dagger}_{i\alpha s}
\vec{\sigma}_{ss^{'}}c_{i\alpha s^{'}}$
and the density operator $n_{i \alpha}=\sum_{s}c^{\dagger}_{i\alpha s}c_{i\alpha s}$,
with $\vec{\sigma}$ representing the
Pauli matrices.
For the hole-doped case, the constraint prohibiting the double-occupancy of the fermion is
\begin{equation}\label{Eq:Constraint1}
\sum_{s} c^\dagger_{i\alpha s} c_{i\alpha s} \leqslant 1.
\end{equation}
For the electron-doped case, we first apply a particle-hole transformation to the $c$-fermions, then enforce the above double-occupancy constraint. The constraint in Eq.~\ref{Eq:Constraint1} can be treated in the standard way by a slave-boson mean-field theory\cite{Kotliar88}.
We introduce a slave boson operator $b_{i\alpha}$ and a fermionic spinon operator $f_{i\alpha s}$ in each site and for each orbital, and rewrite the electron operator as $c_{i\alpha s} = b^\dagger_{i\alpha} f_{i\alpha s}$. The spin operator is rewritten to
$\vec{S}_{i\alpha}=\frac{1}{2}\sum_{s,s^{'}}f^{\dagger}_{i\alpha s}
\vec{\sigma}_{ss^{'}}f_{i\alpha s^{'}}$. In the slave-boson representation, the constraint in Eq.~\ref{Eq:Constraint1} becomes:
\begin{equation}\label{Eq:Constraint2}
 b^\dagger_{i\alpha} b_{i\alpha} + \sum_{s} f^\dagger_{i\alpha s} f_{i\alpha s} = 1.
\end{equation}
In the slave-boson mean-field approach, we introduce a Lagrange multiplier $\lambda$ to impose the constraint on average~\cite{note_Lagrange}, and perform a mean-field decomposition between the slave boson and the fermionic spinon operators. We further assume that the slave bosons are Bose condensed with $\langle b^\dagger_{i\alpha} \rangle = \langle b_{i\alpha} \rangle = \sqrt{|x/2|}$, where $x=\sum_\alpha \langle n_{i\alpha}\rangle-2$, is the doping concentration (with $x<0$ for hole doping and $x>0$ for electron doping). Here we \red{take} into account the fact that in the parent compound, the $d_{xz}$ and $d_{yz}$ orbitals are half-filled, with $n_{xz} + n_{yz}=2$.

With these simplifications, we obtain the following effective Hamiltonian for the spinons:
\begin{eqnarray}\label{Eq:Ham:tJ1J2K_Ren}
 H_{\mathrm{eff}}=&-& \sum_{i<j,\alpha,\beta,s} \tilde{t}_{ij}^{\alpha \beta}
f^{\dagger}_{i\alpha s}f_{j\beta s}+h.c.\nonumber\\
&-&\tilde{\mu} \sum_{i,\alpha}\left[\sum_{s} f^\dagger_{i\alpha s} f_{i\alpha s} -(1-\frac{x}{2})\right] \nonumber \\
&+&\sum_{\langle ij\rangle,\alpha,\beta} J_{1}^{\alpha \beta}
\left(\vec{S}_{i\alpha}\cdot \vec{S}_{j\beta}-\frac{1}{4}n_{i\alpha} n_{j\beta}\right)
\nonumber \\
&+&\sum_{\langle \langle ij\rangle \rangle,\alpha,\beta} J_{2}^{\alpha \beta}
\left(\vec{S}_{i\alpha}\cdot \vec{S}_{j\beta}-\frac{1}{4}n_{i\alpha} n_{j\beta}\right)\nonumber\\
&-&\sum_{\langle ij\rangle} K
\left[(\sum_{\alpha}\vec{S}_{i\alpha})\cdot (\sum_{\beta}\vec{S}_{j\beta})\right]^2,
\end{eqnarray}
where $\tilde{t}_{ij}^{\alpha \beta} = \frac{x}{2} t_{ij}^{\alpha \beta}$, and $\tilde{\mu} = \mu - \lambda$. In Eq.~\ref{Eq:Ham:tJ1J2K_Ren}, the constraint is taken implicitly by renormalizing the hopping matrix with the hole doping concentration. In the following, we work with this effective Hamiltonian, and consider the superconductivity via a BCS mean-field theory.
We define the spin singlet ($\mathcal{S}_{i\alpha,j\beta}$) and triplet ($\mathcal{T}^m_{i\alpha,j\beta}$, $m=0,\pm$) operators in the pairing channel to be
\begin{align}
 \mathcal{S}_{i\alpha,j\beta} &=& f_{i\alpha\downarrow} f_{j\beta\uparrow} - f_{i\alpha\uparrow} f_{j\beta\downarrow}, \\
 \mathcal{T}^0_{i\alpha,j\beta} &=& f_{i\alpha\downarrow} f_{j\beta\uparrow} + f_{i\alpha\uparrow} f_{j\beta\downarrow}, \\
 \mathcal{T}^{\pm}_{i\alpha,j\beta} &=& f_{i\alpha\downarrow} f_{j\beta\downarrow} \pm f_{i\alpha\uparrow} f_{j\beta\uparrow}.
\end{align}
We then rewrite the exchange interactions in Eq.~\ref{Eq:Ham:tJ1J2K_Ren} in terms of $\mathcal{S}_{i\alpha,j\beta}$ and $\mathcal{T}^m_{i\alpha,j\beta}$. For iron pnictides, experiments suggest that the dominant superconducting pairing is in the spin singlet channel. Hence we project the exchange interactions onto the spin singlet sector. We thus obtain
\begin{align}
&& \vec{S}_{i\alpha}\cdot \vec{S}_{j\beta}-\frac{1}{4}n_{i\alpha} n_{j\beta} = -\frac{1}{2} \mathcal{S}^\dagger_{i\alpha,j\beta} \mathcal{S}_{i\alpha,j\beta}, \\
\label{Eq:Ksinglet}&& \left[\left(\sum_{\alpha}\vec{S}_{i\alpha}\right)\cdot \left(\sum_{\beta}\vec{S}_{j\beta}\right)\right]^2 = \frac{9}{32} \sum_{\alpha\beta} \mathcal{S}^\dagger_{i\alpha,j\beta} \mathcal{S}_{i\alpha,j\beta} \nonumber\\
&& +\frac{9}{64} \sum_{\alpha\beta\alpha^\prime\beta^\prime} \mathcal{S}^\dagger_{i\alpha,j\beta} \mathcal{S}^\dagger_{i\alpha^\prime,j\beta^\prime} \mathcal{S}_{i\alpha,j\beta} \mathcal{S}_{i\alpha^\prime,j\beta^\prime}.
\end{align}
In the BCS mean-field theory, the superconductivity corresponds to the Bose condensation of the singlet $\mathcal{S}_{i\alpha,j\beta}$, with the pairing function $\Delta_{\hat{\epsilon}\alpha\beta}=\langle \mathcal{S}_{i\alpha,j\beta} \rangle = \langle \mathcal{S}^\dagger_{i\alpha,j\beta} \rangle$, where $\hat{\epsilon}=i-j$.
We make a further simplification by considering only intra-orbital pairing, namely, $\Delta_{\hat{\epsilon}\alpha\beta} = \Delta_{\hat{\epsilon}\alpha}\delta_{\alpha\beta}$. Including the inter-orbital pairing only changes the results quantitatively. After some algebra, we find the mean-field Hamiltonian in the momentum space:
\begin{align}\label{Eq:Ham:tJ1J2K_mf}
 H_{\mathrm{mf}} &=& \sum_k \psi^\dagger_k
 \left(
   \begin{array}{cc}
     \boldsymbol{\xi}_k - \tilde{\mu} \mathbf{1} & \mathbf{V}_k \\
     \mathbf{V}^\dagger_k & -(\boldsymbol{\xi}_k - \tilde{\mu} \mathbf{1}) \\
   \end{array}
 \right)
 \psi_k + NH_{\mathrm{C}},
\end{align}
where $N$ is the total number of iron sites in the system, and we denote $\psi^T_k=(f_{k1\uparrow}, f_{k2\uparrow}, f^\dagger_{-k1\downarrow},f^\dagger_{-k2\downarrow})$,
\begin{equation}
 \xi^{\alpha\beta}_k=-\sum_{\hat{\epsilon}=\pm \hat{x},\pm \hat{y}, \pm (\hat{x}\pm \hat{y})} \tilde{t}^{\alpha\beta}_{\hat{\epsilon}} \cos(\vec{k}\cdot\hat{\epsilon}).
\end{equation}
and
\begin{eqnarray}
V^{\alpha\beta}_k= & & V_{k\alpha}\delta_{\alpha\beta} \nonumber \\
 V_{k\alpha}= &-& (J_1+\frac{9}{16}K)\sum_{\hat{\epsilon}=\hat{x},\hat{y}} \Delta_{\hat{\epsilon}\alpha} \cos(\vec{k}\cdot\hat{\epsilon}) \nonumber\\
 &-& J_2 \sum_{\hat{\epsilon}=\hat{x}\pm\hat{y}} \Delta_{\hat{\epsilon}\alpha} \cos(\vec{k}\cdot\hat{\epsilon}) \nonumber\\
 &-&  \frac{9}{16}K \sum_{\hat{\epsilon}=\hat{x},\hat{y}} [\sum_{\beta} |\Delta_{\hat{\epsilon}\beta}|^2]\Delta_{\hat{\epsilon}\alpha} \cos(\vec{k}\cdot\hat{\epsilon}).
\end{eqnarray}
The constant term depends on the pairing functions $\Delta_{\hat{\epsilon}\alpha}$ as follows:
\begin{align}
H_{\mathrm{C}} = \left(\frac{J_1}{2}+\frac{9}{32}K\right) \sum_{\alpha,\hat{\epsilon}=\hat{x},\hat{y}} |\Delta_{\hat{\epsilon}\alpha}|^2 + \frac{J_2}{2} \sum_{\alpha,\hat{\epsilon}=\hat{x}\pm\hat{y}} |\Delta_{\hat{\epsilon}\alpha}|^2 \nonumber\\
+ \frac{27}{64}K \sum_{\hat{\epsilon}=\hat{x},\hat{y}} [\sum_{\alpha}|\Delta_{\hat{\epsilon}\alpha}|^2]^2 + \frac{1}{N} \sum_{k\alpha} \xi^{\alpha\alpha}_k - \tilde{\mu} x.
\end{align}
The pairing functions $\Delta_{\hat{\epsilon}\alpha}$ can be determined by minimizing the free energy associated with the mean-field Hamiltonian in Eq.~\ref{Eq:Ham:tJ1J2K_mf}. In the two-orbital model, they can be combined as follows according to the way they transform under the $D_{4h}$ group symmetry operations:
\begin{align}
 s^{A_{1g}}_{x^2+y^2} = (\Delta_{\hat{x}1} + \Delta_{\hat{y}1}) + (\Delta_{\hat{x}2} + \Delta_{\hat{y}2})\\
 s^{B_{1g}}_{x^2+y^2} = (\Delta_{\hat{x}1} + \Delta_{\hat{y}1}) - (\Delta_{\hat{x}2} + \Delta_{\hat{y}2})\\
 d^{B_{1g}}_{x^2-y^2} = (\Delta_{\hat{x}1} - \Delta_{\hat{y}1}) + (\Delta_{\hat{x}2} - \Delta_{\hat{y}2})\\
 d^{A_{1g}}_{x^2-y^2} = (\Delta_{\hat{x}1} - \Delta_{\hat{y}1}) - (\Delta_{\hat{x}2} - \Delta_{\hat{y}2})\\
 s^{A_{1g}}_{x^2y^2} = (\Delta_{\hat{x}+\hat{y}1} + \Delta_{\hat{x}-\hat{y}1}) + (\Delta_{\hat{x}+\hat{y}2} + \Delta_{\hat{x}-\hat{y}2})\\
 s^{B_{1g}}_{x^2y^2} = (\Delta_{\hat{x}+\hat{y}1} + \Delta_{\hat{x}-\hat{y}1}) - (\Delta_{\hat{x}+\hat{y}2} + \Delta_{\hat{x}-\hat{y}2})\\
 d^{B_{2g}}_{xy} = (\Delta_{\hat{x}+\hat{y}1} - \Delta_{\hat{x}-\hat{y}1}) + (\Delta_{\hat{x}+\hat{y}2} - \Delta_{\hat{x}-\hat{y}2})\\
 d^{A_{2g}}_{xy} = (\Delta_{\hat{x}+\hat{y}1} - \Delta_{\hat{x}-\hat{y}1}) - (\Delta_{\hat{x}+\hat{y}2} - \Delta_{\hat{x}-\hat{y}2})
\end{align}
Each of the above eight pairing channels contains an amplitude and a phase.

\begin{figure}[th!]
\centering\includegraphics[
width=80mm]{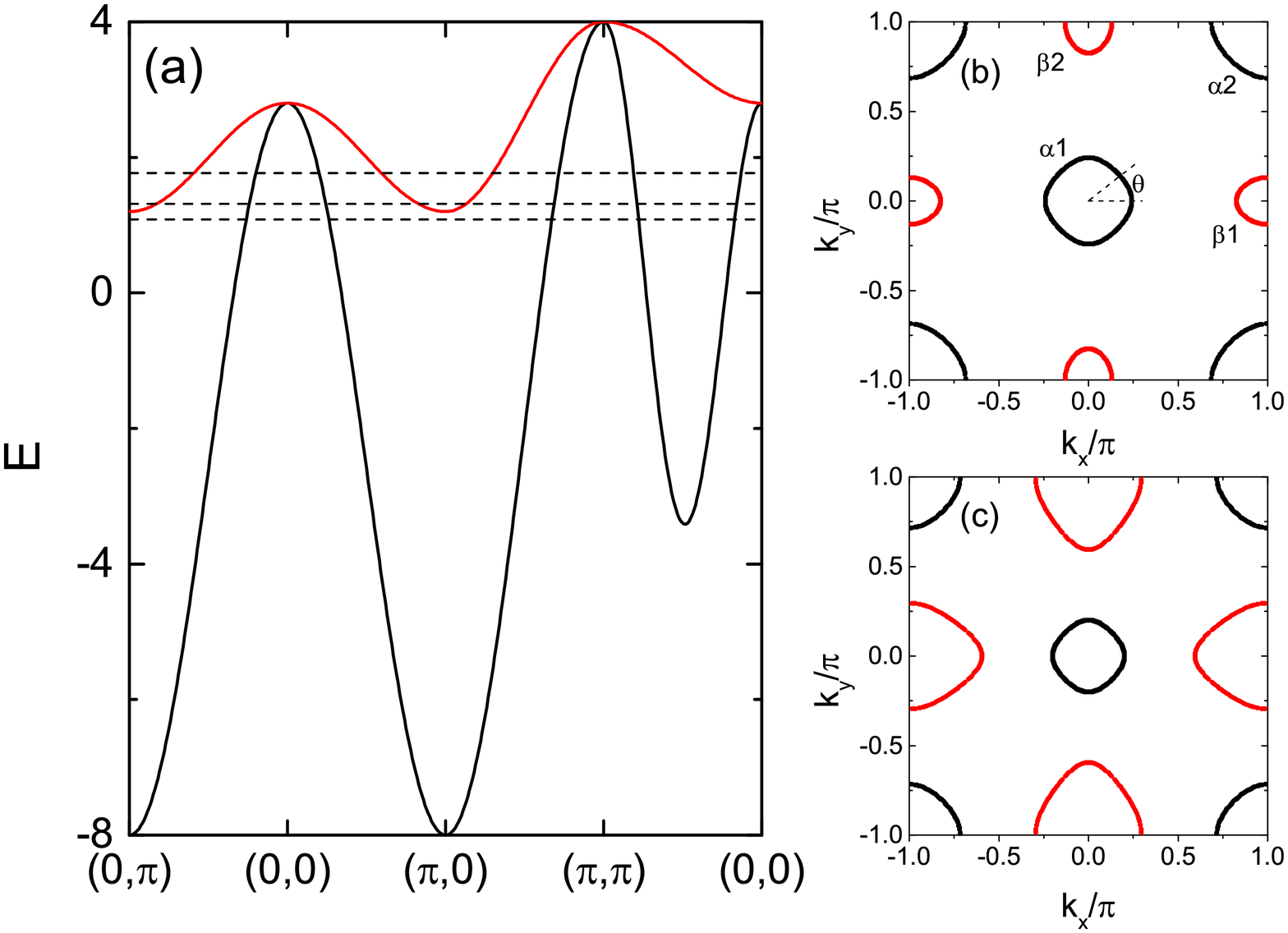}
\caption{(Color online) (a): bandstructure of the two-orbital tight-binding model of Ref~\onlinecite{Raghu}. The dashed lines show the chemical potentials at several electron dopings: $x=-0.32$, $x=-0.16$, $x=0.16$, from bottom to top. (b) and (c): Fermi surface in the 1-Fe Brilluion zone at $x=-0.16$ [in (b)] and $x=0.16$ [in (c)]. In each case, the Fermi surface contains two electron pockets ($\beta1$ and $\beta2$) and two hole pockets ($\alpha1$ and $\alpha2$). $\theta$ is defined as the winding angle along each Fermi pocket.
}
\label{fig:1}
\end{figure}

\begin{figure*}[th!]
\centering\includegraphics[
width=160mm]{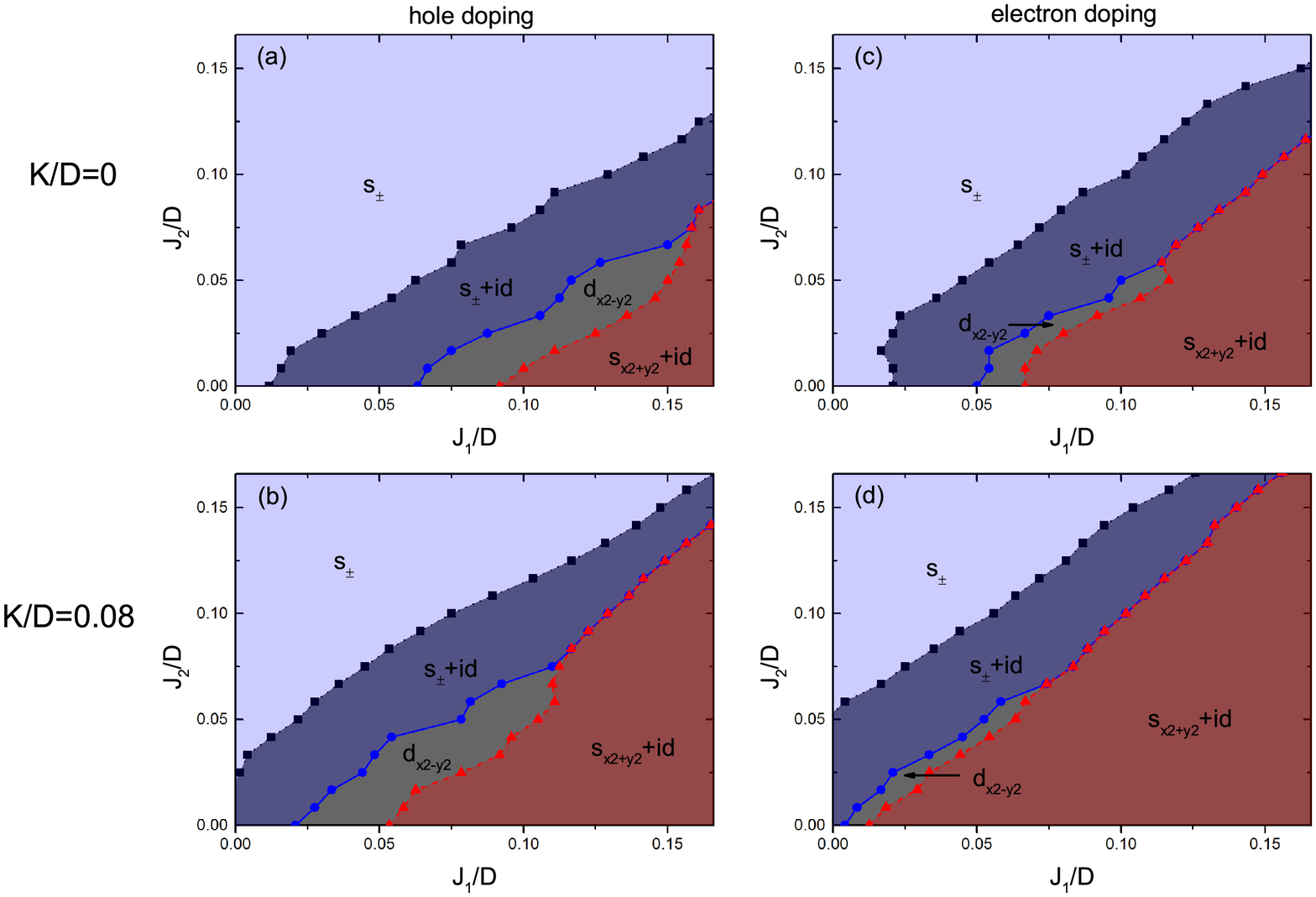}
\vspace{6mm}
\caption{(Color online) Pairing phase diagrams at hole [$x=-0.16$, in (a),(b)] and electron [$x=0.16$, in (c),(d)] dopings without [$K/D=0$, in (a),(c)] and with [$K/D=0.08$, in (b),(d)] the biquadratic interaction. The dominant pairing channel in each phase is, from left to right, $s_{\pm}$ ($s^{A_{1g}}_{x^2y^2}$), $s_{\pm}+id$ ($s^{A_{1g}}_{x^2y^2}+id_{x^2-y^2}^{B_{1g}}$), $d_{x^2-y^2}$ ($d_{x^2-y^2}^{B_{1g}}$), and $s_{x^2+y^2}+id$ ($s^{A_{1g}}_{x^2+y^2}+id_{x^2-y^2}^{B_{1g}}$). The symbols respectively indicate the phase boundaries.
}
\label{fig:2}
\end{figure*}

\section{Mean-field phase diagram}\label{Sec:PhD}

For the fermiology, we take the two-orbital tight-binding model in Ref.~\onlinecite{Raghu}. As shown in Fig.~\ref{fig:1}, within a certain range of electron doping, the Fermi surface contains two hole pockets ($\alpha1$ and $\alpha2$) and two electron pockets ($\beta1$ and $\beta2$). But the bandstructure is highly asymmetric about half-filling ($x=0$), and for large hole doping concentration $x\lesssim-0.25$, the Fermi surface contains only hole pockets. This allows us to study the effects of heavy hole doping.

\begin{figure*}[th!]
\centering\includegraphics[
width=160mm]{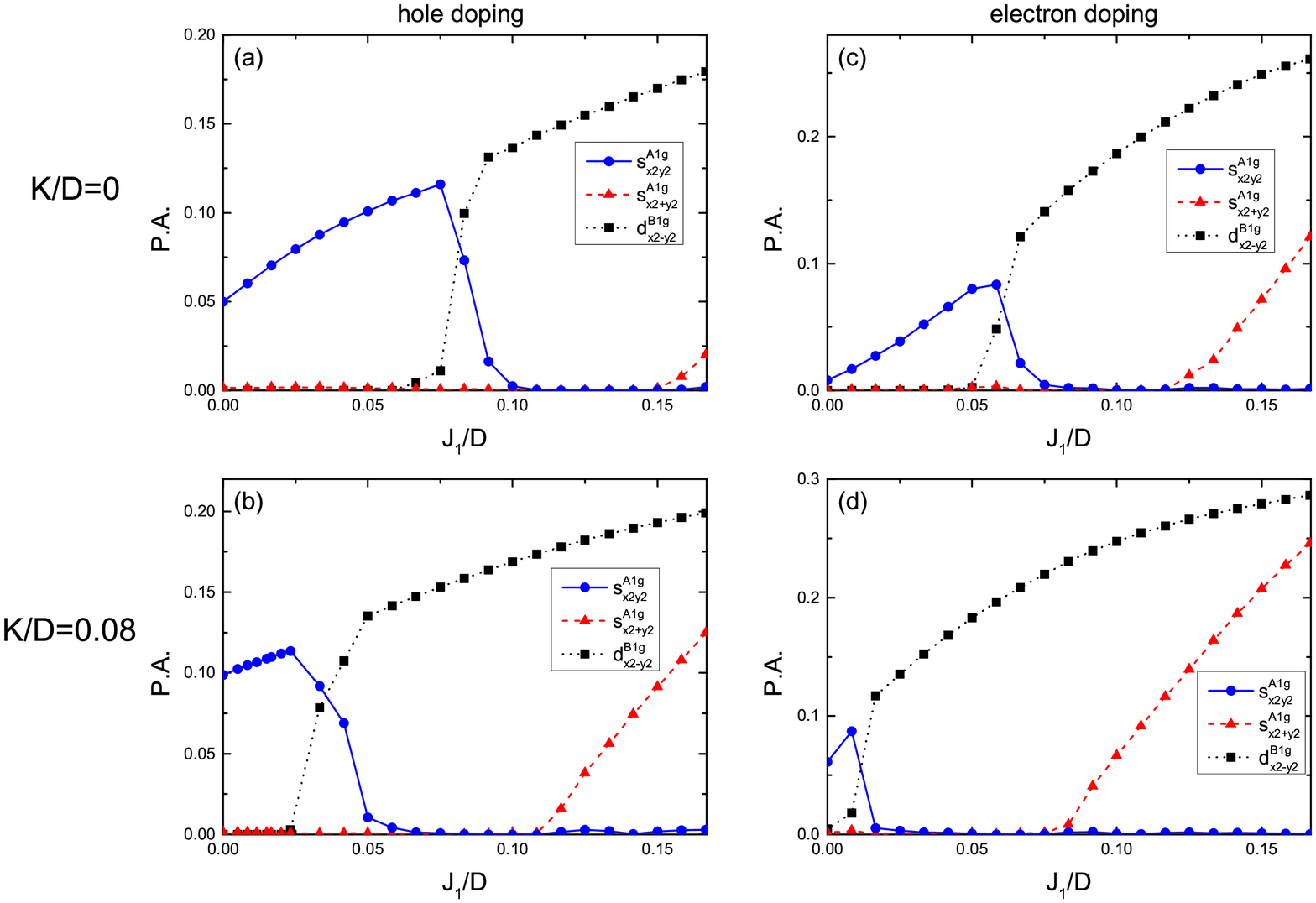}
\caption{(Color online) Pairing amplitudes (P.A.) of the dominant pairing channels at $J_2/D=0.05$ for $x=-0.16$ [in (a),(b)] and $x=0.16$ [in (c),(d)] without [$K/D=0$, in (a),(c)] and with [$K/D=0.08$, in (b),(d)] the biquadratic interaction.
}
\label{fig:3}
\end{figure*}

Figure~\ref{fig:2} compares the superconducting phase diagrams for the $t-J_1-J_2$ (with $K=0$) and $\tjjk$ models at doping concentrations $x=\pm0.16$. The corresponding pairing amplitudes at $J_2/D=0.05$ are shown in Fig.~\ref{fig:3}. Here we have scaled the exchange coupling by the renormalized bandwidth of spinons, $D$. In the $t-J_1-J_2$ model [Fig.~\ref{fig:2}(a) and (c)], for both the hole and electron doping, the phase diagram contains four phases with different dominant pairing symmetries. As clearly shown in Fig.~\ref{fig:2}, when $J_2\gg J_1$, the pairing symmetry is $A_{1g}$, and the leading pairing channel is $s_{x^2y^2}$. The pairing function changes sign between the hole and electron pockets, and is thus denoted as $s_{\pm}$. With increasing $J_1$, the $d_{x^2-y^2}^{B_{1g}}$ channel emerges and coexists with $s_{x^2y^2}^{A_{1g}}$ (see Fig.~\ref{fig:3}). At zero temperature, the phase difference of these two pairing functions is fixed to be $\pi/2$, and this leads
  to a $s_{x^2y^2}^{A_{1g}}+id_{x^2-y^2}^{B_{1g}}$ (denoted as $s_{\pm}+id$) which breaks the time reversal symmetry. Further increasing $J_1$, the pairing amplitudes of the $A_{1g}$ channels vanish, and there is a regime where the pairing symmetry is pure $B_{1g}$, with the dominant pairing to be $d_{x^2-y^2}$. The $A_{1g}$ symmetry reappears at a higher $J_1$ value, but the dominant pairing channel becomes $s_{x^2+y^2}$. The phases between the $A_{1g}$ and $B_{1g}$ are again locked, resulting in an $s_{x^2+y^2}^{A_{1g}}+id_{x^2-y^2}^{B_{1g}}$ pairing (denoted as $s_{x^2+y^2}+id$). The phase boundaries are sensitive to the bandstructure. We see from Fig.~\ref{fig:2} that the $d_{x^2-y^2}$ phase is narrower for the electron doping than for the hole doping. This phase is completely suppressed when the doping concentration changes from $x=0.16$ to $x=0.14$. (See Supplementary Figure 3 of Ref.~\onlinecite{Yu11}.) We discuss the doping effects in more detail in Section~\ref{Sec:Doping}.

\begin{figure}[h!]
\centering\includegraphics[
width=80mm]{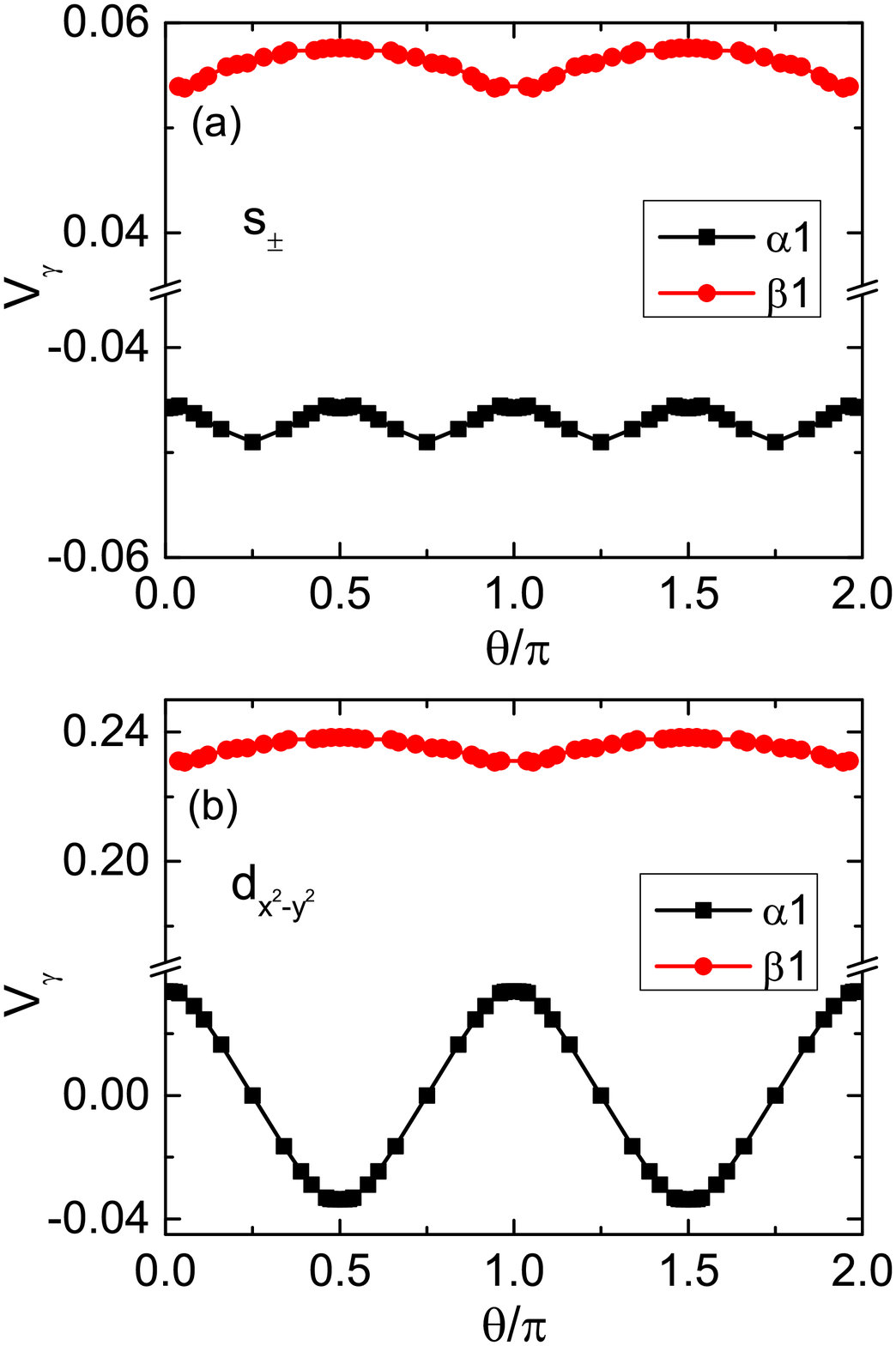}
\caption{(Color online) Angle dependence of pairing function in the band basis, $\tV_{k\gamma}$ along the hole ($\alpha1$) and electron ($\beta1$) Fermi pockets at $x=-0.16$, $K/D=0.08$, and $J_2/D=0.05$. (a): $J_1/D=0.008$, in the $s_{\pm}$ phase; (b): $J_1/D=0.08$, in the $d_{x^2-y^2}$ phase.
}
\label{fig:4}
\end{figure}

\begin{figure}[th!]
\centering\includegraphics[
width=80mm]{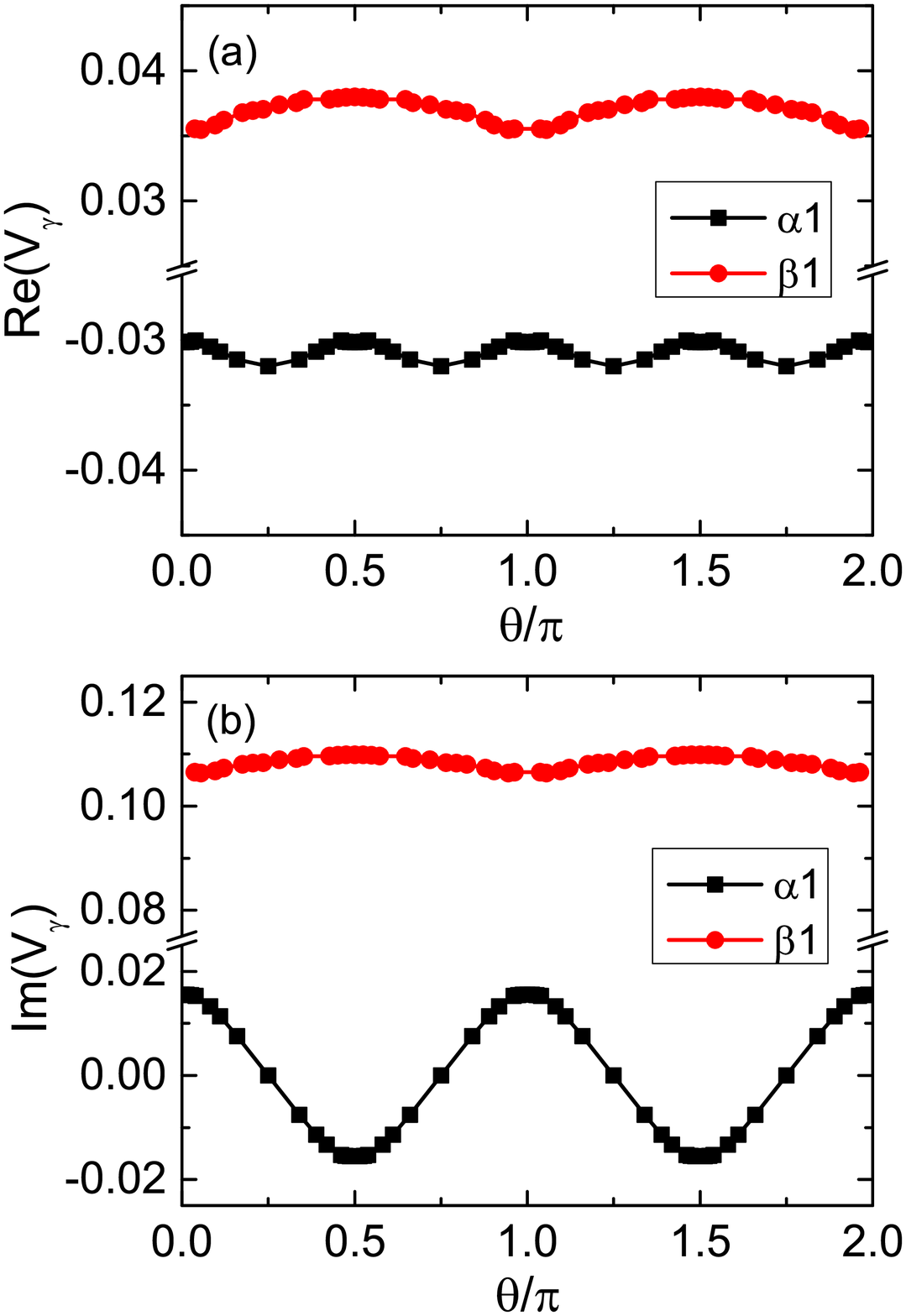}
\caption{(Color online) Angle dependence of the real [in (a)] and imaginary [in (b)] parts of $\tV_{k\gamma}$ along the hole ($\alpha1$) and electron ($\beta1$) Fermi pockets in the $s_{\pm}+id$ phase where $x=-0.16$, $K/D=0.08$, $J_1/D=0.05$, and $J_2/D=0.05$.
}
\label{fig:5}
\end{figure}

Having established  the phase diagram of the $t-J_1-J_2$ model, we now discuss the effects of the biquadratic coupling $K$. By comparing the phase diagrams in Fig.~\ref{fig:2}, we see that when a moderate $K$ is turned on, despite the pairing symmetries remaining the same as in the $t-J_1-J_2$ model, the phase boundaries change a lot. To understand this, note that there are two terms in Eq.~\ref{Eq:Ksinglet}. The first term is bilinear in $\mathcal{S}_{i\alpha,j\beta}$, and can be absorbed into the $J_1$ term. The second term gives nonlinear couplings between the gap functions. If the exchange couplings $J_1$, $J_2$, and $K$ are small compared to the renormalized bandwidth of the spinons, $D$, then $\Delta_{\hat{\epsilon}\alpha\beta}$ is small. As a result, the main effect of the biquadratic term is to renormalize the n.n. exchange coupling $J_1$ by promoting both the $d_{x^2-y^2}^{B_{1g}}$ and the $s_{x^2+y^2}^{A_{1g}}$ pairing channels,
as illustrated in Fig.~\ref{fig:3}. This is the main reason for the change of the phase boundaries. The nonlinearity effect of the gap function appears at sufficiently strong exchange couplings, where several pairing channels become quasi-degenerate. Note that the degeneracy of the pairing channels at infinitely strong $J_1$ (or $J_2$) couplings are associated with gauge symmetries in the exchange interactions,\cite{Goswami10} which are still preserved in presence of the biquadratic coupling. Hence the nonlinearity effects associated with the finite $K$ term will not lead to a strong change of the phase boundaries in the moderate coupling regime, however they may affect the values of pairing amplitudes in each channel.

Superconductivity opens up a gap in the BCS quasiparticle spectrum. In a single-band model, the momentum distribution of this gap directly reflects the superconducting pairing symmetry. In the two-orbital model, we must transform the pairing function $V_{k\alpha}$ into the band basis: $\tV_{k\gamma}=\sum_\alpha|\mathrm{U}_{\alpha\gamma}|V_{k\alpha}$, where the matrix $\mathbf{U}$ diagonalizes $\boldsymbol{\xi}_k$. The $\mathbf{k}$ dependence of $\tV_{k\gamma}$ along the hole ($\alpha1$) and electron ($\beta1$) pockets in the $s_{\pm}$ and $d_{x^2-y^2}$ phases are shown in Fig.~\ref{fig:4}, respectively. In the $s_{\pm}$ phase, the dominant pairing is $s_{x^2y^2}^{A_{1g}}$. The pairing function is nodeless on both electron and hole pockets, but has different signs along these two pockets. We see from Fig.~\ref{fig:4}(a) that $\tV_{k\gamma}$ is isotropic along both the electron and hole pockets, and this leads to an isotropic gap. On the other hand, in the $d_{x^2-y^2}$ phase, where the dominant pairing is $d
 _{x^2-y^2}^{B_{1g}}$, the pairing function has nodes along the hole pockets at $\theta=\pm \pi/4$ (corresponds to $k_x=\pm k_y$). As a result, the gap is highly anisotropic along the hole pockets. Interestingly, in the two 
$s+id$ phases, we find that the gap can be still anisotropic, but it is nodeless. In these phases, nodes appear in either the real or imaginary part of $\tV_{k\gamma}$ (see Fig.~\ref{fig:5}). But due to the different pairing symmetry, the nodes in $\mathrm{Re}(\tV_{k\gamma})$ and $\mathrm{Im}(\tV_{k\gamma})$ necessarily locate at different momenta. The quasiparticle gap is proportional to $\sqrt{[\mathrm{Re}(\tV_{k\gamma})]^2+[\mathrm{Im}(\tV_{k\gamma})]^2}$, and is always nodeless.

\section{Evolution of superconducting states with doping}\label{Sec:Doping}

\begin{figure}[th!]
\begin{center}
\includegraphics[
width=80mm]{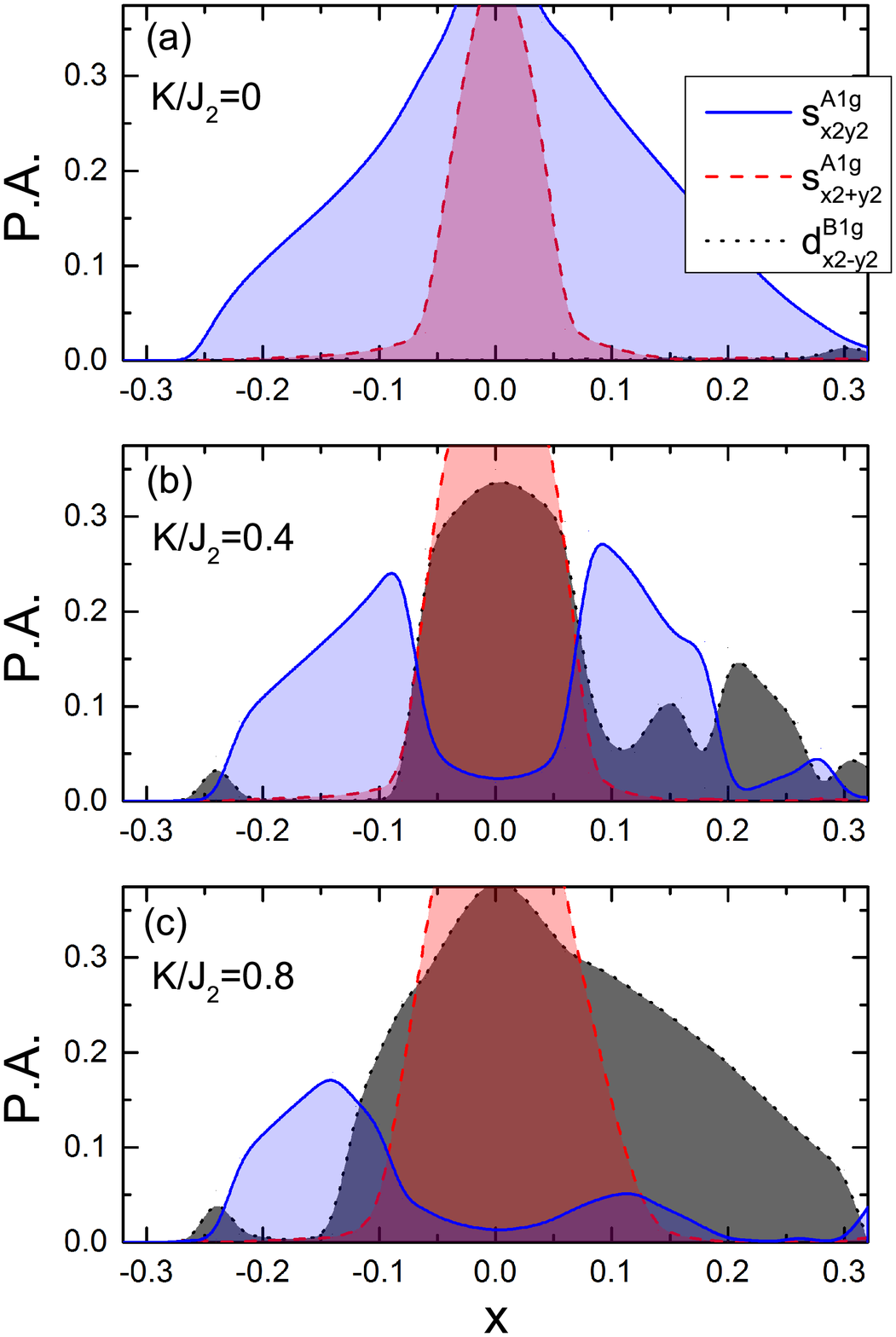}
\end{center}
\caption{(Color online) Doping evolution of the dominant pairing amplitudes (P.A.) for fixed $J_1=J_2$ and various $K/J_2$ values. See text for the choice of $J_1$, $J_2$, and $K$ values.
}
\label{fig:6}
\end{figure}

In the previous section, we have discussed the superconducting phase diagram of the $\tjjk$ model at fixed hole or electron doping. We find that several phases with different pairing symmetries can be stabilized depending on the interactions. An important question with more experimental relevance is that of the pairing symmetry for realistic model parameters, and its doping dependence. In this section we investigate the evolution of superconducting states with doping. In the $\tjjk$ model, the doping concentration $x$ renormalizes the electron bandwidth, $\tilde{t}_{ij}^{\alpha \beta} = \frac{x}{2} t_{ij}^{\alpha \beta}$. To make a fair comparison at different doping concentrations, \red{we take values of the exchange and biquadratic couplings that fit the inelastic neutron scattering data in the parent iron pnictides, and fix these values when changing doping concentration.} For BaFe$_2$As$_2$, the couplings are estimated to be $J_1\approx J_2\approx 20$ meV, and the bare bandwidth projected to the $d_{xz/yz}$ orbitals from DFT calculations is about $3$ eV. From these values, we find $J_1/D=J_2/D\approx0.08$ at optimal (electron/hole) doping $x=\pm0.16$. To study the influence of the biquadratic interaction, we compare the results at three different values of $K$: $K/J_2=0$, $K/J_2=0.4$, and $K/J_2=0.8$. In each case, the dependence of the pairing amplitudes with doping  for the dominant pairing channels is shown in Fig.~\ref{fig:6}. The overall pairing amplitudes are maximal at half-filling, which is understood because  we only consider the superconducting pairing, but have ignored the magnetic order in the model. When $|x|$ is decreased toward half-filling, it
 is expected that the superconductivity is suppressed, and the ground state is eventually antiferromagnetic, as is found experimentally. Hence our results are more pertinent to the optimally and over-doped regimes ($|x|\gtrsim0.1$).

 The pairing amplitudes are strongly suppressed by heavy electron/hole doping. Given that the pairing amplitudes are proportional to the superconducting transition temperature $T_c$, this result is consistent with the very low superconducting $T_c\sim3$ K for the heavily hole doped compound KFe$_2$As$_2$.

From Fig.~\ref{fig:6} we see that the biquadratic interaction may strongly affect the pairing symmetry. For $K=0$, the pairing symmetry is $A_{1g}$ for almost the entire doping regime studied. At any doping concentration, the dominant pairing channel is $s_{\pm}$. This picture changes at a moderate biquadratic coupling $K/J_2=0.4$. Depending on doping, the pairing symmetry is either a pure $A_{1g}$, or a pure $B_{1g}$, or a $A_{1g}+iB_{1g}$. Very interestingly, we find that the dominant pairing is:
\vspace{-1mm}
\begin{enumerate}
\item[(i)] $s_{x^2+y^2}+id$ for $|x|\lesssim0.1$, \vspace{-2mm}
\item[(ii)] $s_{\pm}$ near optimal doping ($0.1\lesssim|x|\lesssim0.2$), and \vspace{-2mm}
\item[(iii)] $d_{x^2-y^2}^{B_{1g}}$ in the overdoped regime ($|x|\gtrsim0.2$).
\end{enumerate}

Such a doping evolution of the pairing symmetry and amplitudes also leads to the change of superconducting gap anisotropy with doping. From the discussion in Sec.~\ref{Sec:PhD} we see that the gap is anisotropic when the $d_{x^2-y^2}^{B_{1g}}$ is dominant. Hence our results suggest a change of pairing from the isotropic $
 s_{\pm}$ in the optimally doped regime to the anisotropic $d_{x^2-y^2}$ in the overdoped regime.

\section{Discussions}\label{Sec:Disc}
As mentioned in Section~\ref{Sec:Model}, besides the degenerate $d_{xz}$ and $d_{yz}$ orbitals, the other three Fe $3d$ orbitals also contribute to the low-energy bandstructure. Therefore generally speaking, a five-orbital model is more appropriate when describing the electronic properties of these materials. For superconducting pairing, several calculations based on five-orbital models have been done.\cite{Kuroki08,Graser,Goswami10,Yu11, Yu14} To see how accurate is the superconducting phase diagram we obtained in Sec.~\ref{Sec:PhD} for the two-orbital model, we compare our results with the strong-coupling phase diagram of a five-orbital $t-J_1-J_2$ model in Ref.~\onlinecite{Goswami10}. We see that the pairing phase diagrams of the two- and five-orbital models are quite similar. This suggests that the simplified two-orbital model already captures the correct pairing symmetry and the dominant pairing channel in each phase, as alluded to earlier in Section~\ref{Sec:Model}. It is a natural result of the strong-coupling theory: the dominant pairing symmetry
 is determined by $J_{1(2)}/D$ and $K/D$, and the details of the Fermi surface structure are secondary. Nonetheless, the orbital character along the Fermi surface is important to the anisotropy of the superconducting gap, especially when the superconducting pairing has a strong orbital selectivity.\cite{Yu14} In this case, the inclusion of other Fe $3d$ orbitals (such as the $d_{xy}$ orbital) may be crucial and will be the subject of future work.

In Sec.~\ref{Sec:Doping} we have shown that the dominant pairing symmetries at different dopings are sensitive to the biquadratic coupling $K$. Without the biquadratic coupling, the dominant pairing is $s_{\pm}$ at any doping. With a moderate $K$, the $s_{\pm}$ pairing is dominant only near optimal doping. For overdoped and underdoped regimes, the dominant pairing channels are respectively $d_{x^2-y^2}$ and $(s_{x^2+y^2}+id_{x^2-y^2})$. The change of pairing symmetry with doping is a consequence of the band renormalization. With increasing $|x|$, the renormalized bandwidth $D$ increases, and $J_{1(2)}/D$ and $K/D$ reduces. The variation of $J_{1(2)}/D$ and $K/D$ accounts for both the change of the pairing symmetry and the dominant pairing amplitude, as we have already seen in Figs.~\ref{fig:2} and \ref{fig:3}. Note that our theory on the doping evolution of superconducting pairing is very different from the weak-coupling theories, in which the doping evolution reflects the interplay between the intra- and inter-pocket electron interactions.\cite{Maitietal11, Chubukov12}

\section{Conclusions}\label{Sec:Conclusion}
In conclusion, we have studied the effect of the biquadratic spin-spin interaction $-K(\vec{S}_i\cdot\vec{S}_j)^2$ on the superconducting states of the iron pnictides via a two-orbital $\tjjk$ model within the slave-boson mean-field theory.
 Unlike  the $t-J_1-J_2$ model, where the dominant pairing channel at any doping concentration is always $s_\pm$  for realistic $J_1/J_2$ ratios, we find that a moderate biquadratic interaction $K$ favors both the $d_{x^2-y^2}^{B_{1g}}$ and $s_{x^2+y^2}^{A_{1g}}$ pairing channels in the $\tjjk$ model, resulting in a serious change of pairing symmetry with doping. Though the dominant pairing is still the isotropic $s_\pm$ (referred to as $s_{x^2y^2}^{A_{1g}}$) near the optimal doping, it changes to $d_{x^2-y^2}^{B_{1g}}$ in the overdoped regime and to $s_{x^2+y^2}^{A_{1g}}+id_{x^2-y^2}^{B_{1g}}$ in the underdoped regime, and in both cases the gap can be anisotropic, as Figs.~\ref{fig:4} and \ref{fig:5} illustrate.

Our results are in qualitative agreement with the gap anisotropy observed in recent ARPES measurements for several iron-based superconductors, including NaFeAs, LiFeAs~\cite{GeFeng13,UmezawaWang12,AllanDavis12} and the heavily hole-doped Ba$_{1-x}$K$_x$Fe$_2$As$_2$.~\cite{Okazaki12,Xu13} Theoretically, the interplay between $s$-wave and $d$-wave pairing channels has been previosuly studied, however to stabilize the $d$-wave state, one had to rely on varying the applied pressure~\cite{Das13},  the pnictogen height~\cite{Platt12}, the $p-d$ orbital hybridization~\cite{Khodas12}, or the strength of N\'eel fluctuations\cite{Fernandes13}. In this study, we show that the $d$-wave state is naturally stabilized in the over-doped regime, without any additional fine-tuning and for realistic values of the biquadratic interaction $K$ inferred~\cite{Yuetal12} from fitting the inelastic neutron spectra to the $\tjjk$ model.

One of the outstanding questions is the effect of electron nematicity on the pairing symmetry and resulting anisotropy of the superconducting order parameter. Recent theoretical work indicates that non-time-reversal symmetry breaking $s_\pm+d_{x^2-y^2}$ pairing can be realized due to coupling to the nematic order parameter~\cite{Livanas12, Fernandes13b}. Study of this effect in the framework of the strong-coupling $\tjjk$ model will be the subject of future work,  potentially relevant to superconductivity in the thin FeSe films~\cite{Song11} and in the stoichiometric single-crystalline FeSe~\cite{Bohmer13}.

\acknowledgements

We thank Q. Si, S.-C. Wang for
useful discussions. This work was supported by the National
Science Foundation of China Grant number 11374361 (R.Y.), and the Welch Foundation grant C-1818 (A.N.)


\begin{thebibliography}{99}
\bibitem{note_Lagrange} Here we have assumed translational symmetry and used the fact that the two orbitals are degenerate.
\end{thebibliography}
\end{document}